\documentclass[prb,twocolumn,showpacs,preprintnumbers,amsmath,amssymb]{revtex4}

\usepackage{graphicx}
\usepackage{dcolumn}
\usepackage{bm}

\begin{document}


\title{Determination of spin Hamiltonian in the Ni$_4$ magnetic molecule}

\author{K.~Iida$^{1,2}$}\email{ki7e@virginia.edu}
\author{S.-H.~Lee$^2$}
\author{T.~Onimaru$^1$}\altaffiliation{Present address: Department of Quantum Matter, Graduate School of Advanced Sciences of Matter, Hiroshima University, Higashi-Hiroshima 739-8530, Japan.}
\author{K.~Matsubayashi$^3$}
\author{T.~J.~Sato$^{1}$}\altaffiliation{Present address: Institute of Multidisciplinary Research for Advanced Materials, Tohoku University, Sendai 980-8577, Japan.}
\affiliation{$^1$Neutron Science Laboratory, Institute for Solid State Physics, University of Tokyo, Kashiwa, Chiba 277-8581, Japan}
\affiliation{$^2$Department of Physics, University of Virginia, Charlottesville, Virginia 22904, USA}
\affiliation{$^3$Division of Physics in Extreme Conditions, Institute for Solid State Physics, University of Tokyo, Kashiwa, Chiba 277-8581, Japan}

\date{\today}

\begin{abstract}
Magnetic excitations in a Ni$_4$ magnetic molecule were investigated by inelastic neutron scattering and bulk susceptibility ($\chi_\text{bulk}$) techniques.
The magnetic excitation spectrum obtained from the inelastic neutron scattering experiments exhibits three modes at energy transfers of $\hbar\omega=0.5$, 1.35, and 1.6~meV.
We show that the energy, momentum, and temperature dependences of the inelastic neutron scattering data and $\chi_\text{bulk}$ can be well reproduced by an effective spin Hamiltonian consisted of intra-molecule exchange interactions, a single-ionic anisotropy, biquadratic interactions, and Zeeman term.
Under a hydrostatic pressure, the bulk magnetization decreases with increasing pressure, which along with the biquadratic term indicates spin-lattice coupling present in this system.
\end{abstract}

\maketitle

\section{Introduction}
Magnetic molecules~\cite{MolecularMagnet,SSM} provide us an excellent opportunity to study quantum behaviors in magnetism, because their effective spin Hamiltonians can be exactly diagonalized and compared to experimental data.
Previous studies on magnetic molecules such as Mn$_{12}$ and Fe$_8$ revealed superparamagnetic behaviors and tunneling between quantum mechanical states.\cite{Thomas,Fe8_1}
These phenomena were explained by treating these molecules as isolated entities with strong single-axis anisotropy~\cite{Mn12} and transverse fourth-order term.\cite{Mn12_3}
It was also found that the V$_{15}$ cluster exhibits a hysteresis loop with dissipative spin reversal in pulsed field magnetization measurements, which was explained by Landau-Zener transition and phonon-bottleneck effect.\cite{V15}

\begin{figure}[b]
\includegraphics[width=6.833cm, height=4.1cm]{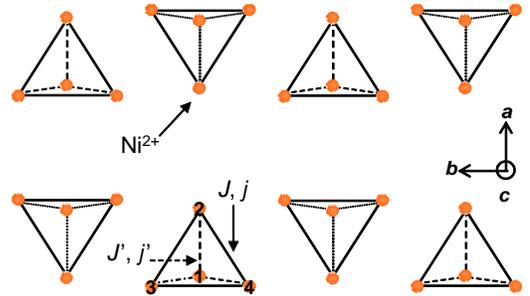}
\caption{\label{Fig:Structure}
(Color online)
Schematic view of the crystal structure of the Ni$_4$ magnetic molecule.
Solid circles represent Ni$^{2+}$ ions.
Solid and dashed lines represent the exchange and biquadratic interactions, and their parameters, $J$, $J'$, $j$, and $j'$, are defined in Eq.~(\ref{Eq:Hamiltonian}).
}
\end{figure}

Ni$_4$ is another intriguing molecular magnet because the antiferromagnetic Ni$^{2+}$ ($s=1$) ions form a tetrahedron,\cite{Ni4_1} which may lead to geometrically frustration.
The full chemical formula of the Ni$_4$ cluster is $[$Mo$_{12}$O$_{30}$($\mu_2$-OH)$_{10}$H$_2$\{Ni(H$_2$O)$_3$\}$_4$$]$$\cdot$14H$_2$O, and its crystal structure is shown in Fig.~\ref{Fig:Structure}; four Ni$^{2+}$ ions form a slightly distorted tetrahedron, and tetrahedra directing oppositely in the $c$-axis are arranged alternately.
Distances between Ni$^{2+}$ ions within a cluster are 6.69, 6.70, 6.62, and 6.60~\AA, whereas the shortest distance between Ni$^{2+}$ ions that belong to different clusters is 7.15~\AA.
Previous bulk property measurements using bulk susceptibility, high field magnetization, electron paramagnetic resonance, optical conductivity, and magneto-optical response~\cite{Ni4_1, Ni4_2, Ni4_3} showed that the dominant interaction in the system is antiferromagnetic, which is due to the superexchange interaction through the Ni-O-Mo-O-Ni bonds.\cite{Ni4_2}
The most interesting property of the Ni$_4$ nanaomagnet is an adiabatic change with non-equidistant steps observed in the magnetization measurements as a function of external magnetic field.\cite{Ni4_2,Ni4_3,Ni4_5}
This was explained by a model Hamiltonian that consists of a field-dependent exchange parameters, a single-ion anisotropy, and a biquadratic interaction.\cite{Ni4_3}
The spin Hamiltonian should be determined by a more direct tool than the bulk property measurements to check its validity.

We have performed inelastic neutron scattering measurements on a powder sample of the Ni$_4$ molecule to investigate the energy ($\hbar\omega$), momentum ($Q$), and temperature ($T$) dependences of the magnetic excitations.
Discrete excited levels were observed at 0.5, 1.35, and 1.6~meV, and $Q$-dependences of each mode have peaks at 0.6 and 1.6~\AA$^{-1}$.
By analyzing the inelastic neutron scattering data, we determine the spin effective Hamiltonian of the spin Ni$_4$ cluster without an external field; a model Hamiltonian consists of intra-molecule exchange interaction, single-ion anisotropy, and biquadratic interaction.\cite{Ni4_3}
The existence of the biquadratic interactions suggests strong spin-lattice coupling in the system, which is consistent with the suppression of the bulk magnetism by an application of a hydrostatic pressure.

\section{Experimental Details}
A 4.5~g deuterated powder sample was prepared using the procedure described in Ref.~\onlinecite{Ni4_1}.
Our prompt-gamma neutron activation analysis showed that about 50\% hydrogen was substituted by deuterium.
A small amount of the sample of 200~mg was used for bulk magnetization using a SQUID magnetometer at the ambient pressure as well as under the hydrostatic pressure up to 0.92~GPa.
For the pressure experiment, 20~mg sample was put into a teflon cell which was then filled with Daphne oil and was set to a piston-cylinder device.\cite{Pressure}
A reference sample of Sn was also put in the cell, and the transition temperature of Sn was used to determine the hydrostatic pressure.\cite{Sn}
Background for the pressure experiments was measured using the empty pressure device, and subtracted from the data.

The remaining 4.3~g sample was used for two sets of neutron scattering experiments.
The first set of the measurements was performed on the cold-neutron triple-axis spectrometer SPINS at the NIST Center for Neutron Research.
A vertically focusing pyrolytic graphite (PG) monochromator and a horizontal focusing PG analyzer were used to increase the sensitivity of the measurements. 
Energy of the scattered neutrons was fixed to be $E_{f}=3.0$~meV, resulting in an instrumental resolution of 117~$\mu$eV (FWHM, or full width at half maximum) at the elastic position.
Energy of the incident neutrons was changed to measure the scattering intensity as a function of energy transfer, $\hbar\omega$.
The energy resolution at $\hbar\omega=0.5$, 1.35, or 1.6~meV is estimated to be 141, 189, or 212~$\mu$eV (FWHM), respectively.\cite{ResolutionFunction}
Higher order contaminations were eliminated using a cooled Be filter placed after the sample.
The second set of experiments was performed at the cold neutron triple-axis spectrometer HER at the JRR-3M research reactor with $E_{f}=5.0$~meV.
A vertically focusing monochromator and a double-focusing (i.e., both horizontal and vertical focusing) analyzer were employed to increase the sensitivity.
The nonmagnetic background was measured at 30~K and subtracted from the low temperature data to obtain the magnetic scattering intensity, $I(\mathbf{Q},\hbar\omega)$,
\begin{eqnarray}
I(\mathbf{Q},\hbar\omega)&=&\sum_{\alpha, \beta}\sum_{a,b}\sum_{i,f}\left(\delta_{\alpha,\beta}-Q^\alpha Q^\beta/Q^2\right)F^2(Q)\nonumber\\
&\times&p_{i}<i|S_a^\alpha e^{-i\mathbf{Q}\cdot\mathbf{r}_a}|f><f|S_b^\beta e^{i\mathbf{Q}\cdot\mathbf{r}_b}|i>\label{Eq:SQw}\\
&\times&\delta(E_{i}-E_{f}+\hbar\omega)\nonumber
\end{eqnarray}
where $|i>$ ($|f>$) is the initial (final) eigenstate, $p_i$ is the Boltzmann factor for the state $|i>$ ($p_i=n_ie^{-E_i/k_\text{B}T}/\sum_jn_je^{-E_j/k_\text{B}T}$ where $n_i$ represents degeneracy of $|i>$), $E_i$ ($E_f$) is the initial (final) energy of the system, $\mathbf{S}_{a,b}$ and $\mathbf{r}_{a,b}$ are the spin operator and position of the Ni$^{2+}$ ions at site $a$, $b$ in the molecule, respectively, $\alpha$ and $\beta$ represents the vector component of $\mathbf{S}$, and $F(Q)$ is the magnetic form factor of the Ni$^{2+}$ ion.\cite{MFfactor}

\section{Results and discussion}

\begin{figure}[t]
\includegraphics[width=8.4cm, height=9.1cm]{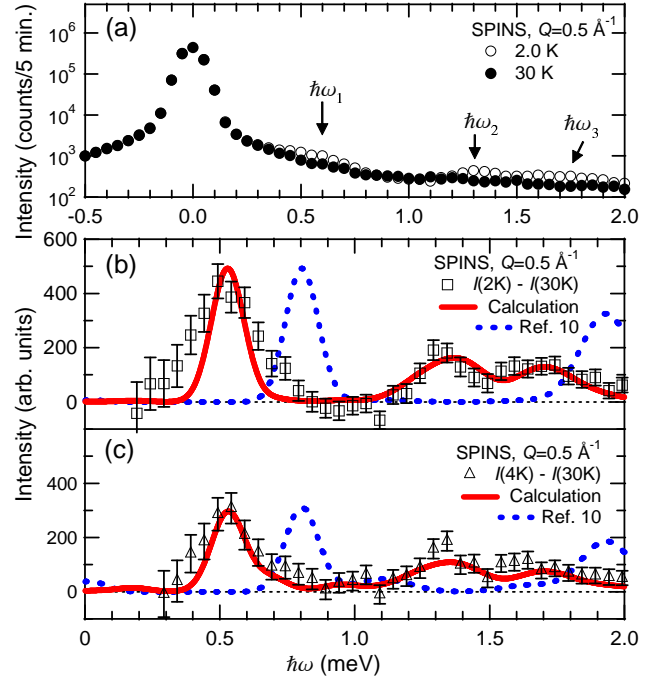}
\caption{\label{Fig:INS}
(Color online)
(a) The $\hbar\omega$-dependences, $I(\hbar\omega)$, measured at $Q=0.5$~\AA$^{-1}$ and $T=2.0$ and 30~K.
The $\hbar\omega$-dependences of the magnetic intensities at (b) 2.0~K obtained by $I(\hbar\omega,2.0\text{K})-I(\hbar\omega,30\text{K})$ and at (c) 4.0~K by $I(\hbar\omega,4.0\text{K})-I(\hbar\omega,30\text{K})$, respectively.
Solid and dashed lines are described in the main text.
}
\end{figure}

\begin{figure}[t]
\includegraphics[width=8.4cm]{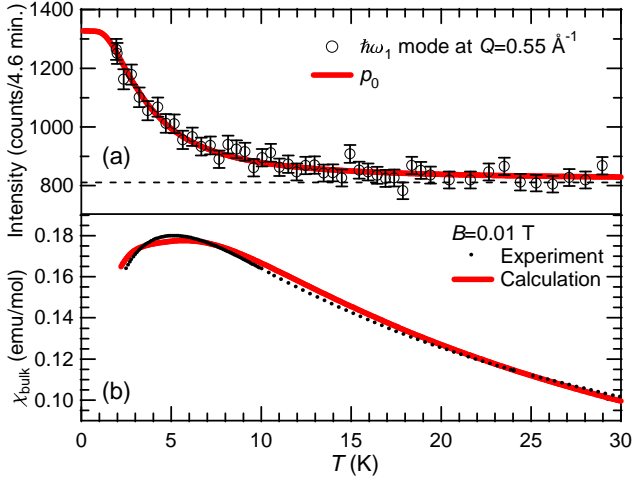}
\caption{\label{Fig:Sus}
(Color online)
(a) $T$-dependence of the $\hbar\omega_1=0.5$~meV mode measured at $Q=0.55$~\AA$^{-1}$.
Solid line is the Boltzmann factor for the ground state, $p_0$ explained in the main text.
Dashed line represents background.
(b) Experimental (dots) and model calculated (line) $T$-dependence of the bulk susceptibility, $\chi_\text{bulk}$, under an external magnetic field of $B=0.01$~T.
}
\end{figure}

Figure~\ref{Fig:INS}(a) shows the $\hbar\omega$-dependence of the neutron scattering intensities measured at $Q=0.5$~\AA$^{-1}$ at two different temperatures of $T=2.0$ and 30~K.
The strong sharp peak centered at $\hbar\omega=0$~meV is mainly due to the incoherent scattering from hydrogen in the sample. 
At $T=2.0$~K, in addition to the strong incoherent scattering, three distinct inelastic excitations exist centered at around $\hbar\omega_1=0.5$, $\hbar\omega_2=1.35$, and $\hbar\omega_3=1.6$~meV.
At $T=30$~K, the three inelastic peaks broaden and become indistinguishable with the incoherent scattering.
We take the $T=30$~K data as the background and subtract it from the $T=2$~K data to obtain the magnetic energy spectrum.
The result, $I(2\text{K})-I(30\text{K})$, is plotted in Fig.~\ref{Fig:INS}(b).
In order to investigate the temperature dependence, we have performed the same measurements at 4.0~K.
$I(4\text{K})-I(30\text{K})$ measured at $Q=0.5$~\AA$^{-1}$ is shown in Fig.~\ref{Fig:INS}(c).
The intensities of the three peaks become weaker at 4.0~K, indicating that these three peaks are magnetic.
In order to study $T$-dependence further, we have measured $T$-dependence of the $\hbar\omega_1$ mode and bulk susceptibility ($\chi_\text{balk}$).
As Fig.~\ref{Fig:Sus} shows, upon cooling the $\hbar\omega_1$ mode slowly appears at low temperatures and rapidly increases below 5~K, which coincides with the downturn in $\chi_\text{balk}$.
This supports that the excitations observed in the neutron scattering spectra are due to a development of antiferromagnetic correlations.

\begin{figure}[t]
\includegraphics[width=7.7cm]{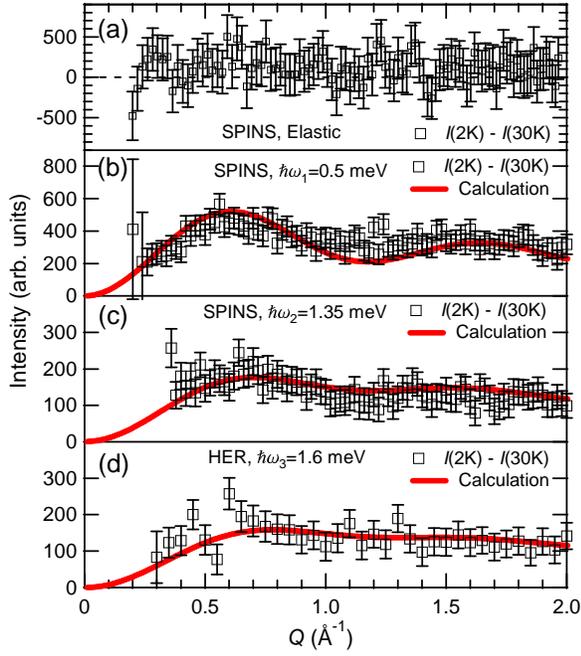}
\caption{\label{Fig:Q}
(Color online)
$Q$-dependences of (a) elastic, (b) $\hbar\omega_1=0.5$~meV, (c) $\hbar\omega_2=1.35$~meV, and (d) $\hbar\omega_3=1.6$~meV modes.
Solid lines are explained in the text.
}
\end{figure}

We have also measured the $Q$-dependences of the three excitations at $T=2.0$ or 0.7~K.
As shown in Figs.~\ref{Fig:Q}(b)--\ref{Fig:Q}(d), the 0.5, 1.35, and 1.6~meV excitations are peaked at $Q_0=0.6$ and 1.6~\AA$^{-1}$, while no clear magnetic intensity is observed in the $Q$-dependence at the elastic channel [Fig.~\ref{Fig:Q}(a)].
All three excitations weaken at high $Q$, confirming that they are magnetic in origin.
The $Q$-dependence of the three excitations are similar to each other, indicating that they have the same origin.
The $Q$-dependence of the intensity has the information about what is the magnetic entity in real space, which will be discussed later in detail.

Following the previous study,\cite{Ni4_3} we assume that the spin Hamiltonian of the Ni$_4$ spin cluster consists of the exchange interaction, the single ion anisotropy, the biquadratic interaction, and the Zeeman term;
\begin{eqnarray}
\mathcal{H}=&-&\sum\limits_{i\neq j=1}^4J_{ij}\mathbf{S}_i\cdot\mathbf{S}_j+D\sum\limits_{i=1}^4(\mathbf{e}_i\cdot\mathbf{S}_i)^2\label{Eq:Hamiltonian}\\
&-&\sum\limits_{i\neq j=1}^4j_{ij}(\mathbf{S}_i\cdot\mathbf{S}_j)^2+g\mu_\text{B}\mathbf{B}\cdot\sum\limits_{i=1}^4\mathbf{S}_i,\nonumber
\end{eqnarray}
where $J_{ij}$, $D$, and $j_{ij}$ are the parameters of the exchange interaction, the single ion anisotropy, and the biquadratic interaction, respectively.
By considering the crystallographic symmetry in the Ni$_4$ molecule,\cite{Ni4_3} there are two different values for $J_{ij}$ and $j_{ij}$ as described by the solid and dashed lines in Fig.~\ref{Fig:Structure}, and we define $J$, $J'$, $j$ and $j'$ using $J_{ij}$ and $j_{ij}$ as $J=J_{12}=J_{13}=J_{14}$, $J'=J'_{23}=J'_{24}=J'_{34}$, $j=j_{12}=j_{13}=j_{14}$, and $j'=j'_{23}=j'_{24}=j'_{34}$.
$g$ and $\mu_\text{B}$ represent the geometric factor and the Bohr magneton.
$\mathbf{e}_i$ describes a local anisotropic axis; considering the geometrical frustration, $\mathbf{e}_i$ points radially outward from the center of the tetrahedron.\cite{Ni4_3}

\begin{figure}[t]
\includegraphics[width=8.54cm]{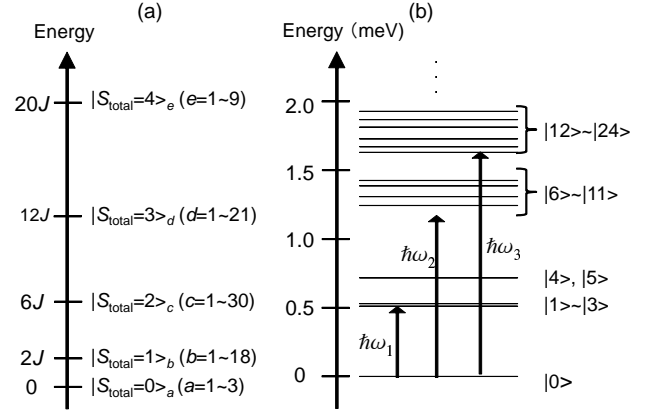}
\caption{\label{Fig:Energy}
(a) Energy levels and eigenstates of the Ni$_4$ magnetic molecule for $\mathcal{H}_0=-J\sum \mathbf{S}_i\cdot\mathbf{S}_j$.
The subscript, $a\sim e$, represents the degeneracy of each state.
Some eigenstates are described in Table~\ref{Tab:Eigenvector1}.
(b) Some of the energy levels and eigenstates of the Ni$_4$ spin cluster with $\mathcal{H}$ written in Eqs.~(\ref{Eq:Hamiltonian}) and (\ref{Eq:Parameters}).
Some eigenstates $|n>$ are described in Table~\ref{Tab:Eigenvector2}.
}
\end{figure}

\begin{table*}[t]
\begin{center}
\caption{
Some of low energy eigenstates of the Ni$_4$ magnetic molecule assuming it has the simple hamiltonian $\mathcal{H}_0$ described in Fig.~\ref{Fig:Energy}(a).
The eigenstates are described by the linear combinations of the $z$-components of the four Ni$^{2+}$ ($s=1$) spins $|s_1^zs_2^zs_3^zs_4^z>$.
$\uparrow$, $0$, and $\downarrow$ represent $s^z =$ 1, 0, and -1, respectively.
For simplicity, the terms whose coefficients are less than 0.2 were not written here for the linear combinations, but they were included in our calculations.
}\label{Tab:Eigenvector1}
\begin{tabular}{c|l}
\hline
$|S_\text{total}=2>_{30}$ & $-0.25|\uparrow0\downarrow\downarrow>+0.26|0\downarrow\uparrow\downarrow>-0.28|\downarrow000>+0.26|0000>-0.21|\uparrow\uparrow00>+0.22|0\downarrow\downarrow\uparrow>+0.4|\uparrow\uparrow\downarrow\uparrow>$\\
\hline
$|S_\text{total}=2>_{29}$ & $0.23|\uparrow\uparrow\downarrow\downarrow>-0.25|00\uparrow\downarrow>-0.25|00\downarrow\uparrow>+0.24|\uparrow00\uparrow>-0.22|0\uparrow0\uparrow>+0.23|\downarrow\downarrow\uparrow\uparrow>-0.29|\uparrow\downarrow\uparrow\uparrow>$\\
\hline
$|S_\text{total}=2>_{28}$ & $0.22|0\downarrow\uparrow\downarrow>-0.22|\uparrow\downarrow\uparrow\downarrow>-0.3|0\uparrow\downarrow0>+0.24|0000>-0.3|0\downarrow\uparrow0>-0.2|\uparrow0\uparrow0>$\\
 & $-0.22|\downarrow\uparrow\downarrow\uparrow>+0.27|\uparrow\downarrow\uparrow\uparrow>$\\
\hline
$|S_\text{total}=2>_{27}$ & $-0.22|0\uparrow0\downarrow>+0.22|\uparrow\uparrow0\downarrow>+0.25|\uparrow\downarrow00>+0.25|\downarrow\uparrow00>-0.27|00\uparrow0>-0.22|0\downarrow0\uparrow>$\\
 & $+0.26|\uparrow\downarrow0\uparrow>$\\
\hline
$|S_\text{total}=2>_{26}$ & $-0.22|00\downarrow\downarrow>+0.27|\uparrow\uparrow\uparrow\downarrow>+0.28|\uparrow\downarrow\uparrow0>-0.22|\uparrow0\downarrow\uparrow>+0.23|\uparrow\downarrow\uparrow\uparrow>$\\
\hline
$|S_\text{total}=2>_{24}$ & $0.23|00\downarrow\downarrow>-0.27|0\uparrow\downarrow\downarrow>+0.21|32>+0.22|\uparrow\downarrow00>+0.22|\downarrow\uparrow00>-0.27|\uparrow0\uparrow0>$\\
 & $-0.21|\downarrow\downarrow0\uparrow>+0.21|\uparrow\downarrow\uparrow\uparrow>$\\
\hline
$|S_\text{total}=2>_{23}$ & $0.23|00\downarrow\downarrow>+0.24|\uparrow0\downarrow\downarrow>-0.23|\downarrow\uparrow\downarrow\downarrow>+0.35|\downarrow\downarrow\uparrow\downarrow>-0.23|\uparrow\uparrow\uparrow\downarrow>+0.2|\downarrow\downarrow\downarrow\uparrow>$\\
\hline
$|S_\text{total}=2>_{21}$ & $0.4|\uparrow\downarrow\downarrow\downarrow>-0.26|\downarrow00\downarrow>-0.32|0\downarrow\downarrow0>+0.24|\downarrow0\downarrow0>+0.27|\downarrow\downarrow00>$\\
\hline
$|S_\text{total}=2>_{20}$ & $0.24|\downarrow00\downarrow>-0.27|\downarrow\downarrow\uparrow\downarrow>+0.29|0\downarrow\uparrow\downarrow>+0.23|\downarrow0\uparrow\downarrow>-0.26|\downarrow0\downarrow0>-0.23|0\downarrow00>$\\
 & $0.22|\downarrow\downarrow\downarrow\uparrow>+0.23|\downarrow0\downarrow\uparrow>-0.35|\uparrow\uparrow\downarrow\uparrow>$\\
\hline
$|S_\text{total}=2>_{19}$ & $-0.22|\uparrow0\downarrow\downarrow>-0.38|\downarrow\uparrow\downarrow\downarrow>+0.23|\downarrow00\downarrow>+0.32|\downarrow0\downarrow0>-0.37|\downarrow\downarrow00>$\\
\hline
$|S_\text{total}=2>_{18}$ & $0.27|\downarrow00\downarrow>-0.21|0\downarrow\uparrow\downarrow>-0.21|\downarrow0\uparrow\downarrow>-0.28|\downarrow0\downarrow0>-0.2|0\uparrow\downarrow0>+0.21|0\downarrow00>$\\
 & $-0.2|0\downarrow\uparrow0>+0.46|\downarrow\downarrow\downarrow\uparrow>-0.21|\downarrow0\downarrow\uparrow>$\\
\hline
$|S_\text{total}=2>_{17}$ & $-0.3|0\uparrow\downarrow\downarrow>+0.2|\uparrow00\downarrow>+0.23|\downarrow0\uparrow\downarrow>+0.22|\uparrow\uparrow\uparrow\downarrow>+0.35|\uparrow\downarrow\downarrow0>-0.21|0\downarrow00>$\\
 & $+0.25|\uparrow\uparrow\downarrow\uparrow>+0.2|\downarrow00\uparrow>$\\
\hline
$|S_\text{total}=1>_{18}$ & $-0.2|\downarrow0\uparrow\downarrow>+0.25|\downarrow\uparrow\downarrow0>-0.31|0\downarrow00>+0.47|0\downarrow\downarrow\uparrow>-0.45|\downarrow0\downarrow\uparrow>$\\
\hline
$|S_\text{total}=1>_{17}$ & $-0.21|\uparrow\downarrow0\downarrow>+0.36|0\downarrow\uparrow\downarrow>-0.26|00\uparrow\downarrow>+0.2|\downarrow000>-0.33|\downarrow\downarrow\uparrow0>+0.26|00\downarrow\uparrow>$\\
\hline
$|S_\text{total}=1>_{15}$ & $-0.29|44>-0.22|\uparrow\downarrow\uparrow0>+0.3|\downarrow\uparrow\uparrow0>-0.21|\uparrow\downarrow0\uparrow>+0.31|\downarrow\uparrow0\uparrow>+0.43|0\downarrow\uparrow\uparrow>$\\
\hline
$|S_\text{total}=1>_{14}$ & $-0.33|\uparrow\downarrow\uparrow\downarrow>+0.27|\uparrow0\uparrow\downarrow>+0.22|\downarrow\uparrow\uparrow\downarrow>+0.21|\uparrow0\downarrow0>-0.28|\uparrow\downarrow\uparrow0>-0.21|\downarrow0\uparrow0>$\\
 & $-0.22|\uparrow\downarrow\downarrow\uparrow>-0.24|\uparrow0\downarrow\uparrow>+0.33|\downarrow\uparrow\downarrow\uparrow>+0.3|\uparrow\downarrow0\uparrow>$\\
\hline
$|S_\text{total}=1>_{13}$ & $-0.31|\uparrow0\downarrow\downarrow>+0.23|\uparrow\downarrow0\downarrow>+0.21|16>-0.21|\uparrow\downarrow\uparrow\downarrow>-0.27|\downarrow0\uparrow\downarrow>-0.25|0\downarrow00>$\\
 & $+0.43|\downarrow\downarrow\uparrow0>+0.21|\downarrow\uparrow\downarrow\uparrow>$\\
\hline
$|S_\text{total}=1>_{12}$ & $-0.33|\uparrow\uparrow\downarrow\downarrow>+0.24|0\uparrow0\downarrow>-0.22|\uparrow0\uparrow\downarrow>+0.33|\uparrow0\downarrow0>-0.33|\downarrow0\uparrow0>-0.25|\downarrow\uparrow\uparrow0>$\\
 & $+0.23|\uparrow0\downarrow\uparrow>-0.24|0\downarrow0\uparrow>+0.23|\downarrow\uparrow0\uparrow>+0.33|\downarrow\downarrow\uparrow\uparrow>$\\
\hline
$|S_\text{total}=1>_{11}$ & $0.34|0\uparrow\downarrow\downarrow>+0.22|0\uparrow0\downarrow>-0.21|\downarrow\uparrow\uparrow\downarrow>-0.26|00\downarrow0>-0.21|\downarrow\uparrow\downarrow0>+0.21|\uparrow\downarrow00>$\\
 & $+0.21|\downarrow000>-0.21|\downarrow\uparrow00>+0.21|\uparrow\downarrow\downarrow\uparrow>+0.26|\downarrow0\downarrow\uparrow>-0.34|\downarrow\downarrow0\uparrow>-0.22|0\downarrow0\uparrow>$\\
\hline
$|S_\text{total}=1>_{10}$ & $-0.28|\uparrow00\downarrow>+0.25|\uparrow\downarrow\uparrow\downarrow>-0.25|\downarrow\uparrow\downarrow\uparrow>+0.2|\uparrow\downarrow0\uparrow>+0.28|\downarrow00\uparrow>$\\
\hline
$|S_\text{total}=1>_{9}$ & $0.25|0\downarrow\uparrow\downarrow>-0.32|\downarrow0\uparrow\downarrow>+0.21|\uparrow0\uparrow\downarrow>+0.24|\uparrow\downarrow\downarrow0>-0.21|\downarrow\uparrow\downarrow0>-0.21|00\uparrow0>$\\
\hline
$|S_\text{total}=1>_{8}$ & $0.28|\uparrow\uparrow\downarrow\downarrow>-0.27|\uparrow\downarrow\uparrow\downarrow>+0.22|00\uparrow\downarrow>-0.32|\uparrow0\uparrow\downarrow>+0.24|0\uparrow\uparrow\downarrow>+0.2|\uparrow\downarrow00>$\\
 & $-0.2|\downarrow\uparrow00>+0.23|\uparrow\downarrow\uparrow0>-0.22|00\downarrow\uparrow>+0.27|\downarrow\uparrow\downarrow\uparrow>-0.28|\downarrow\downarrow\uparrow\uparrow>$\\
\hline
$|S_\text{total}=1>_{7}$ & $-0.27|0\uparrow\downarrow\downarrow>-0.26|\uparrow\uparrow\downarrow\downarrow>-0.22|\uparrow\downarrow0\downarrow>+0.23|000\downarrow>+0.23|\uparrow\downarrow\downarrow0>+0.28|0\uparrow\downarrow0>$\\
 & $-0.23|\downarrow000>-0.28|0\downarrow\uparrow0>-0.23|0\downarrow\downarrow\uparrow>+0.22|0\uparrow\downarrow\uparrow>+0.25|\downarrow\downarrow0\uparrow>+0.26|\downarrow\downarrow\uparrow\uparrow>$\\
\hline
$|S_\text{total}=1>_{6}$ & $-0.26|\uparrow\downarrow0\downarrow>-0.21|00\uparrow\downarrow>-0.35|0\uparrow\downarrow0>+0.35|0\downarrow\uparrow0>+0.21|00\downarrow\uparrow>+0.2|000\uparrow>$\\
\hline
$|S_\text{total}=1>_{5}$ & $-0.31|\uparrow0\downarrow\downarrow>+0.31|000\downarrow>-0.34|0\downarrow\uparrow\downarrow>+0.27|\downarrow0\uparrow\downarrow>+0.26|\uparrow\downarrow\downarrow>+0.3|\downarrow\uparrow\downarrow>$\\
\hline
$|S_\text{total}=1>_{4}$ & $0.51|\uparrow\uparrow0\downarrow>-0.39|0\uparrow\uparrow\downarrow>-0.43|\uparrow\uparrow\downarrow0>+0.24|00\uparrow0>+0.32|0\uparrow\downarrow\uparrow>$\\
\hline
$|S_\text{total}=1>_{3}$ & $-0.24|\uparrow\uparrow0\downarrow>-0.35|\uparrow\uparrow0\downarrow>+0.44|\uparrow000>-0.25|\uparrow\downarrow\uparrow0>-0.27|00\uparrow0>+0.29|0\uparrow\downarrow\uparrow>$\\
 & $-0.31|000\uparrow>+0.44|0\downarrow\uparrow\uparrow>$\\
\hline
$|S_\text{total}=1>_{2}$ & $0.28|0\uparrow0\downarrow>-0.31|\downarrow\uparrow\uparrow\downarrow>+0.21|00\downarrow0>+0.31|\uparrow\downarrow\downarrow\uparrow>+0.3|\downarrow\downarrow0\uparrow>-0.28|0\downarrow0\uparrow>$\\
\hline
$|S_\text{total}=1>_{1}$ & $0.33|\uparrow\downarrow0\downarrow>-0.25|000\downarrow>+0.22|\downarrow0\uparrow\downarrow>-0.22|00\uparrow\downarrow>+0.23|\downarrow\uparrow\uparrow\downarrow>-0.21|\uparrow\downarrow\downarrow0>+0.28|00\downarrow0>$\\
 & $-0.22|\downarrow\uparrow\downarrow0>-0.23|\uparrow\downarrow\downarrow\uparrow>+0.22|00\downarrow\uparrow>$\\
\hline
$|S_\text{total}=0>_{3}$ & $-0.22|\uparrow00\downarrow>-0.22|0\uparrow0\downarrow>-0.22|\uparrow0\downarrow0>-0.22|0\uparrow\downarrow0>+0.3|\uparrow\downarrow00>-0.22|0\downarrow\uparrow0>$\\
 & $-0.22|\downarrow0\uparrow0>-0.22|0\downarrow0\uparrow>-0.22|\downarrow00\uparrow>+0.45|\downarrow\uparrow\downarrow\uparrow>$\\
\hline
$|S_\text{total}=0>_{2}$ & $-0.31|\uparrow\downarrow\uparrow\downarrow>+0.33|00\uparrow\downarrow>-0.35|\downarrow\uparrow\uparrow\downarrow>+0.33|\uparrow\downarrow00>-0.33|0000>+0.33|\downarrow\uparrow>-0.35|\uparrow\downarrow\downarrow\uparrow>$\\
 & $+0.33|00\downarrow\uparrow>-0.31|\downarrow\uparrow\downarrow\uparrow>$\\
\hline
$|S_\text{total}=0>_{1}$ & $0.29|\uparrow00\downarrow>-0.29|0\uparrow0\downarrow>-0.31|\uparrow\downarrow\uparrow\downarrow>+0.27|\downarrow\uparrow\uparrow\downarrow>-0.29|\uparrow0\downarrow0>+0.29|0\uparrow\downarrow0>+0.29|0\downarrow\uparrow0>$\\
 & $-0.29|\downarrow0\uparrow0>+0.27|\uparrow\downarrow\downarrow\uparrow>-0.31|\downarrow\uparrow\downarrow\uparrow>-0.29|\downarrow00\uparrow>+0.29|\downarrow00\uparrow>$\\
\hline
\end{tabular}
\end{center}
\end{table*}

\begin{table*}[t]
\begin{center}
\caption{
Some of low-energy eigenstates and energies of the Ni$_4$ magnetic molecule for the Hamiltonian $\mathcal{H}$ of Eqs.~(\ref{Eq:Hamiltonian}) and (\ref{Eq:Parameters}).
The eigenstates of $\mathcal{H}$ can be described by linear combinations of the states of $\mathcal{H}_0$ shown in Table~\ref{Tab:Eigenvector1}.
For simplicity, the terms whose coefficients have an absolute value less than 0.3 were not written here for the linear combinations, but they were included in our calculations.}\label{Tab:Eigenvector2}
\begin{tabular}{c|l}
\hline
$E_n-E_0$ (meV) &\ \ \ \ \ \ \ \ \ \ \ \ \ \ \ \ \ \ \ \ \ \ \ \ \ \ \ \ \ \ \ \ \ \ \ \ \ \ \ \ \ \ $|n>$\\
\hline
\vdots &\ \ \ \ \ \ \ \ \ \ \ \ \ \ \ \ \ \ \ \ \ \ \ \ \ \ \ \ \ \ \ \ \ \ \ \ \ \ \ \ \ \ \ \ \vdots\\
\hline
1.930 & $|24>\simeq(-0.76+0.02i)|S_\text{total}=1>_{17}+(-0.4-0.03i)|S_\text{total}=1>_{18}$\\
\hline
1.930 & $|23>\simeq(-0.08+0.4i)|S_\text{total}=1>_{17}+(0.22-0.73i)|S_\text{total}=1>_{18}$\\
\hline
1.869 & $|22>\simeq(0.65i)|S_\text{total}=2>_{17}+(-0.02-0.3i)|S_\text{total}=2>_{29}$\\
\hline
1.869 & $|21>\simeq(-0.01+0.33i)|S_\text{total}=2>_{18}+(-0.02-0.35i)|S_\text{total}=2>_{21}+(0.01+0.33i)|S_\text{total}=2>_{26}$\\
 & $\ \ \ \ \ \ \ +(0.02-0.5i)|S_\text{total}=2>_{27}+(0.03-0.32i)|S_\text{total}=2>_{28}$\\
\hline
1.862 & $|20>\simeq(0.53-0.23i)|S_\text{total}=2>_{19}+(-0.31+0.14i)|S_\text{total}=2>_{21}+(-0.32+0.14i)|S_\text{total}=2>_{23}$\\
 & $\ \ \ \ \ \ \ +(0.34-0.16i)|S_\text{total}=2>_{28}$\\
\hline
1.814 & $|19>\simeq(-0.49-0.01i)|S_\text{total}=2>_{21}+(0.63+0.02i)|S_\text{total}=2>_{23}+(-0.31+0.05i)|S_\text{total}=2>_{24}$\\
\hline
1.814 & $|18>\simeq(0.36-0.08i)|S_\text{total}=2>_{17}+(0.3-0.1i)|S_\text{total}=2>_{19}+(0.31-0.11i)|S_\text{total}=2>_{20}$\\
 & $\ \ \ \ \ \ \ +(0.36-0.07i)|S_\text{total}=2>_{24}+(-0.49+0.13i)|S_\text{total}=2>_{30}$\\
\hline
1.796 & $|17>\simeq(-0.46+0.15i)|S_\text{total}=1>_{1}+(0.2+0.31i)|S_\text{total}=1>_{3}+(0.04-0.37i)|S_\text{total}=1>_{4}$\\
 & $\ \ \ \ \ \ \ +(-0.3+0.31i)|S_\text{total}=1>_{5}+(0.34+0.34i)|S_\text{total}=1>_{6}$\\
\hline
1.726 & $|16>\simeq(-0.21+0.31i)|S_\text{total}=1>_{1}+(0.5+0.39i)|S_\text{total}=1>_{2}+(-0.23+0.32i)|S_\text{total}=1>_{3}$\\
\hline
1.726 & $|15>\simeq(0.19+0.57i)|S_\text{total}=1>_{2}+(0.37-0.24i)|S_\text{total}=1>_{3}+(0.3-0.08i)|S_\text{total}=1>_{4}$\\
\hline
1.669 & $|14>\simeq(-0.32+0.56i)|S_\text{total}=1>_{5}+(-0.55-0.31i)|S_\text{total}=1>_{6}$\\
\hline
1.669 & $|13>\simeq(-0.49+0.34i)|S_\text{total}=1>_{1}+(-0.04-0.42i)|S_\text{total}=1>_{3}+(0.42+0.34i)|S_\text{total}=1>_{4}$\\
\hline
1.632 & $|12>\simeq(0.21+0.35i)|S_\text{total}=1>_{3}+(0.44-0.26i)|S_\text{total}=1>_{4}+(0.44-0.02i)|S_\text{total}=1>_{5}$\\
 & $\ \ \ \ \ \ \ +(-0.03-0.49i)|S_\text{total}=1>_{6}$\\
\hline
1.424 & $|11>\simeq(-0.16-0.48i)|S_\text{total}=1>_{8}+(-0.47+0.2i)|S_\text{total}=1>_{9}+(-0.3+0.04i)|S_\text{total}=1>_{10}$\\
 & $\ \ \ \ \ \ \ +(0.12+0.37i)|S_\text{total}=1>_{14}$\\
\hline
1.383 & $|10>\simeq(-0.4-0.09i)|S_\text{total}=1>_{7}+(-0.16+0.51i)|S_\text{total}=1>_{8}+(-0.29+0.24i)|S_\text{total}=1>_{10}$\\
 & $\ \ \ \ \ \ \ +(-0.37-0.12i)|S_\text{total}=1>_{13}$\\
\hline
1.383 & $|9> \simeq(0.33-0.47i)|S_\text{total}=1>_{9}+(0.34+0.1i)|S_\text{total}=1>_{10}+(0.32+0.18i)|S_\text{total}=1>_{14}$\\
 & $\ \ \ \ \ \ \ +(0.34+0.18i)|S_\text{total}=1>_{15}$\\
\hline
1.306 & $|8> \simeq(0.37-0.15i)|S_\text{total}=1>_{7}+(-0.42+0.03i)|S_\text{total}=1>_{9}+(0.32+0.33i)|S_\text{total}=1>_{10}$\\
 & $\ \ \ \ \ \ \ +(-0.2+0.25i)|S_\text{total}=1>_{11}$\\
\hline
1.243 & $|7> \simeq(0.53+0.07i)|S_\text{total}=1>_{12}+(-0.05+0.52i)|S_\text{total}=1>_{13}+(0.25+0.28i)|S_\text{total}=1>_{14}$\\
 & $\ \ \ \ \ \ \ +(-0.29+0.18i)|S_\text{total}=1>_{15}$\\
\hline
1.243 & $|6> \simeq(0.31-0.43i)|S_\text{total}=1>_{12}+(-0.47-0.23i)|S_\text{total}=1>_{13}+(-0.31+0.17i)|S_\text{total}=1>_{15}$\\
\hline
0.714 & $|5> \simeq(0.55-0.06i)|S_\text{total}=0>_1+(-0.49+0.07i)|S_\text{total}=0>_2+(-0.02-0.65i)|S_\text{total}=0>_3$\\
\hline
0.714 & $|4> \simeq(0.36-0.33i)|S_\text{total}=0>_1+(-0.33+0.29i)|S_\text{total}=0>_2+(0.44+0.6i)|S_\text{total}=0>_3$\\
\hline
0.531 & $|3> \simeq(-0.61i)|S_\text{total}=1>_{7}+(-0.34i)|S_\text{total}=1>_{8}+(0.02+0.57i)|S_\text{total}=1>_{10}$\\
\hline
0.531 & $|2> \simeq(0.78+0.26i)|S_\text{total}=1>_{11}$\\
\hline
0.512 & $|1> \simeq(0.28+0.5i)|S_\text{total}=1>_{12}+(-0.26-0.47i)|S_\text{total}=1>_{14}+(0.27+0.49i)|S_\text{total}=1>_{15}$\\
\hline
    0 & $|0> \simeq(-0.33-0.58i)|S_\text{total}=0>_1+(-0.37-0.65i)|S_\text{total}=0>_2$\\
\hline
\end{tabular}
\end{center}
\end{table*}

If the Ni$_4$ cluster has uniform exchange couplings only, the spin Hamiltonian will be simply written as $\mathcal{H}_0=-J\sum\mathbf{S}_i\cdot\mathbf{S}_j$.
The ground state of $\mathcal{H}_0$ is a triply degenerate state with zero total spin, $|S_\text{total}=0>_{1\sim3}$.
All the eigenstates of $\mathcal{H}_0$ can be easily calculated, some of which at low energies are illustrated in Fig.~\ref{Fig:Energy}(a) and listed in Table~\ref{Tab:Eigenvector1}.
For the general $\mathcal{H}$ with non-uniform exchange $J_{ij}$, single-anisotropy $D$, and biquadratic $j_{ij}$, the eigenstates can be written as linear combinations of the eigenstates of $\mathcal{H}_0$.
For instance, the ground state of $\mathcal{H}$, denoted by $|0>$, becomes a singlet state that is a linear combination of the triply degenerate ground state $|S_\text{total}=0>_{1\sim3}$ of $\mathcal{H}_0$.
The other two linear combinations of $|S_\text{total}=0>_{1\sim3}$ gain energy for $\mathcal{H}$.
These states and other low-energy excited states, and their energies were obtained by the exact diagonalization of $\mathcal{H}$ for several different sets of values for the parameters $J$, $J'$, $D$, $j$, and $j'$.
The optimum parameters were determined by comparing both the calculated energies and intensities of the allowed transitions between the states with the observed energies and intensities of the excitation modes, $\hbar\omega_1$, $\hbar\omega_2$, and $\hbar\omega_3$;
\begin{eqnarray}
J/k_\text{B} &=&-3.69(3)\ \text{K},\nonumber\\
J'/k_\text{B}&=&-3.19(2)\ \text{K},\nonumber\\
D/k_\text{B} &=&-2.47(2)\ \text{K},\label{Eq:Parameters}\\
j/k_\text{B} &=&-0.11(1)\ \text{K},\nonumber\\
j'/k_\text{B}&=&\ \ 1.52(1)\ \text{K}.\nonumber
\end{eqnarray}
This result clearly show the biquadratic interactions present in the Ni$_4$ cluster.
The resulting low-energy eigenstates of $\mathcal{H}$ for the optimal parameters are listed in Table.~\ref{Tab:Eigenvector2}.
The $\hbar\omega$-dependence of neutron scattering intensities for the three low-energy excitations were calculated by Eq.~(\ref{Eq:SQw}), averaged for powder, and convoluted with Gaussians.
The fitting results shown as solid lines in Figs.~\ref{Fig:INS}(b) and \ref{Fig:INS}(c) reproduce our data well.
The $Q$-dependences of the excitations can be also reproduced by the model as shown in Figs.~\ref{Fig:Q}(b)--\ref{Fig:Q}(d).
On the other hand, the calculated $\hbar\omega$-dependences using the previous parameters reported in Ref.~\onlinecite{Ni4_3} ($J/k_\text{B}=-3.2$~K, $J'/k_\text{B}=-3.1$~K, $D/k_\text{B}=-1.0$~K, $j/k_\text{B}=1.6$~K, and $j'/k_\text{B}=0$~K) cannot fit the experimental $\hbar\omega$-dependences at all [see the dashed lines in Figs.~\ref{Fig:INS}(b) and \ref{Fig:INS}(c)].

$T$-dependence of the $\hbar\omega_1$ mode can be reproduced very well by the Boltzmann factor for the ground state, $p_0=1/\sum_jn_je^{-E_j/k_\text{B}T}$, as described by solid line in Fig.~\ref{Fig:Sus}(a).
$T$-dependence of $\chi_\text{bulk}$ can also be reproduced well by our model as shown in Fig.~\ref{Fig:Sus}(b).
In the calculation of $\chi_\text{balk}$, the geometrical factor was taken to be $g=2.22$.\cite{Ni4_1,Ni4_2,Ni4_3,Ni4_5}
The background due to isolated Ni$^{2+}$ ions was estimated by the Curie law and added to the calculated $\chi_\text{bulk}$.

\begin{figure}[t]
\includegraphics[width=8.4cm, height=6.3cm]{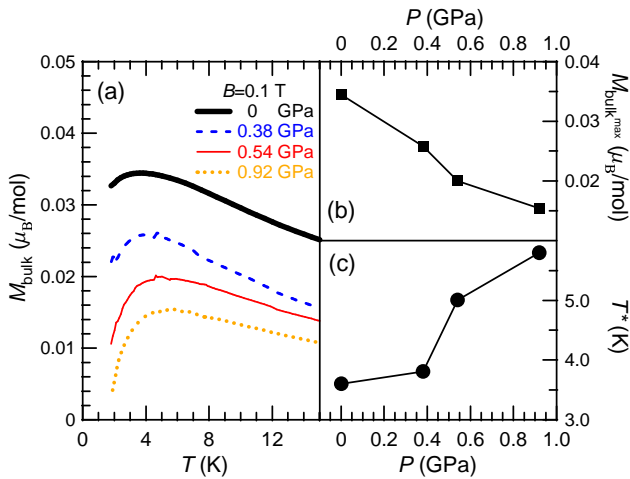}
\caption{\label{Fig:Pressure}
(Color online)
$T$-dependence of the bulk magnetization, $M_\text{bulk}$, measured with $B=0.1$~T under a hydrostatic pressure of $P=0$, 0.38, 0.54, and 0.92~GPa.
$P$-dependences of (b) the maximum value of $M_\text{bulk}$ ($M_\text{bulk}^\text{max}$) and of (c) the peak temperature ($T^*$).
}
\end{figure}

The existence of the biquadratic terms in $\cal H$ indicates a spin-lattice coupling in this system.\cite{biquadratic1}
In order to confirm this, we have measured the bulk magnetization, $M_\text{bulk}$, under a hydrostatic pressure ($P$).
Figure~\ref{Fig:Pressure}(a) shows $M_\text{bulk}$ obtained with $B = 0.1$ T under a hydrostatic pressure of $P=0$, 0.38, 0.54, and 0.92~GPa.
As the pressure increased, the maximum value of $M_\text{bulk}$ ($M_\text{bulk}^\text{max}$) decreases [Fig.~\ref{Fig:Pressure}(b)], while the peak temperature ($T^*$) increases [Fig.~\ref{Fig:Pressure}(c)].
These results represent that $P$ enhances antiferromagnetic correlations by shortening the distance between the Ni$^{2+}$ ions, which may explain the existence of the biquadratic term in this system.

\section{Conclusion}
We have determined the effective spin Hamiltonian, $\mathcal{H}$, in the deuterated Ni$_4$ magnetic molecule by using inelastic neutron scattering and exact diagonalization techniques.
$\hbar\omega$-, $Q$-, and $T$-dependences of neutron scattering intensities due to the low-$\hbar\omega$ excitations centered at 0.5, 1.35, and 1.6~meV as well as $T$-dependence of the bulk susceptibility can be well accounted by $\mathcal{H}$ consisted of exchange interaction, single-ion anisotropy, biquadratic interaction, and Zeeman term.

\begin{acknowledgments}
K.I. acknowledges Global COE Program ``the Physical Sciences Frontier", MEXT, Japan.
We also acknowledge the financial support from the US-Japan Cooperative Program on Neutron Scattering.
Work at UVa was supported by the U.S. Department of Energy, Office of Basic Energy Sciences, Division of Materials Sciences and Engineering under DE-FG02-10ER46384.
\end{acknowledgments}

\end{document}